\documentclass[reprint,
superscriptaddress,frontmatterverbose,showpacs,preprintnumbers,showkeys,nofootinbib,
nobibnotes,amsmath,amssymb,pra,floatfix,twocolumn
]{revtex4-1}
\usepackage{graphics}
\usepackage{ae}
\usepackage[T1]{fontenc} 
\usepackage[english]{babel}
\usepackage[utf8]{inputenc}
\usepackage{amsmath}
\usepackage{graphicx}
\usepackage{color}
\usepackage{dcolumn}
\usepackage{bm}
\usepackage{hyperref}
\usepackage[mathlines]{lineno}
\usepackage{xspace}
\usepackage{algorithm}
\usepackage{algorithmic}
\usepackage{subfigure}
\usepackage{amsfonts}
\usepackage{amsmath}
\usepackage{enumerate}
\usepackage{framed}
\usepackage{bm}
\usepackage{amsthm}
\input{epsf}

\renewcommand{\eqref}[1]{{Eq.~(\ref{#1})}}

\usepackage[utf8]{inputenc}
\usepackage{amsmath}
\usepackage{amsthm}
\usepackage{eucal}
\usepackage{amssymb}
\usepackage{mathrsfs}
\usepackage{physics}
\usepackage{graphicx} 
\begin{document}

\title{Modulational Instability of the time-fractional Ivancevic option pricing model and the Coupled Nonlinear volatility and option price model}
\author{C. Gaafele}
	\email {christopher.gaafele@studentmail.biust.ac.bw}
	\affiliation{Department of Physics and Astronomy, Botswana International University of Science and Technology, Private Mail Bag 16 Palapye, Botswana}

\author{Edmond B. Madimabe}
	\email {me21100032@studentmail.biust.ac.bw}
	\affiliation{Department of Physics and Astronomy, Botswana International University of Science and Technology, Private Mail Bag 16 Palapye, Botswana}

\author{K. Ndebele}
	\email {karabo.ndebele@studentmail.biust.ac.bw}
	\affiliation{Department of Physics and Astronomy, Botswana International University of Science and Technology, Private Mail Bag 16 Palapye, Botswana}

 \author{P. Otlaadisa}
	\email {otlaadisap@biust.ac.bw}
	\affiliation{Department of Physics and Astronomy, Botswana International University of Science and Technology, Private Mail Bag 16 Palapye, Botswana}

 \author{B. Mozola}
	\email {brantony.mozola@studentmail.biust.ac.bw}
	\affiliation{Department of Physics and Astronomy, Botswana International University of Science and Technology, Private Mail Bag 16 Palapye, Botswana}

\author{T. Matabana}
	\email {thabang.matabana@studentmail.biust.ac.bw}
	\affiliation{Department of Physics and Astronomy, Botswana International University of Science and Technology, Private Mail Bag 16 Palapye, Botswana}

 \author{K. Seamolo}
	\email {so21100033@studentmail.biust.ac.bw}
	\affiliation{Department of Physics and Astronomy, Botswana International University of Science and Technology, Private Mail Bag 16 Palapye, Botswana}
\author{P. Pilane}
	\email {pp18001207@studentmail.biust.ac.bw}
	\affiliation{Department of Physics and Astronomy, Botswana International University of Science and Technology, Private Mail Bag 16 Palapye, Botswana}
\date{May 2024}

\begin{abstract}
We study the time-fractional Ivancevic option pricing model and the coupled nonlinear volatility and option price model via both modulational instability (MI) analysis and direct simulations. For the coupled volatility and option pricing model the coupling term for both the volatility and the option price equation is the same, the MI results are dependent on it, and the stability of the volatility exists for the same condition as that of the price. The numerical simulations are done to confirm the conditions of MI. For the time-fractional model the analysis shows that for some values of the Hurst exponent MI exists for negative
values of the adaptive market heat potential. Also, the sign of the volatility does not affect the MI,
even though for some values of the volatility the MI can be suppressed. Direct numerical simulation
shows the existance of solitons for negative values of the adaptive market potential where instabilty
exists due to the value of the Hurst exponent.
\end{abstract}
\maketitle

\section{Introduction}
\label{sec1}
Nowadays, the field of economics and finance is extensively researched globally. Experts and researchers use various technological and scientific tools to develop sophisticated products aimed at optimizing daily life situations. As users engage with these devices, they strive to maximize benefits and achieve optimal outcomes. Traditionally, users rely on their existing knowledge to assess product values, leading to various challenges for buyers, sellers, online platforms, and banks both locally and internationally. These issues are systematically explored using scientific principles, resulting in innovative approaches that highlight the importance of monitoring financial markets. The intricate modeling of global financial markets generates comprehensive information systems, particularly dynamic systems that enable detailed analysis of production processes. Observing the transmission of products from producers to consumers through various channels like highways, air travel, and shipping is also crucial. Initial steps in this process involve formulating mathematical models, whether complex or real-valued, incorporating wave functions. Numerous models have been developed to extract wave distributions, helping to understand current trends and future directions. Soliton theory is often used due to its ability to provide precise information about periodic, singular, dark, bright, complex, and traveling dynamics. These insights are essential for comprehending, predicting, controlling, and planning for complex production behaviors. Notably, researchers like Sharp et al. have investigated stochastic differential equations (SDE) in finance~\cite{sharp1990stochastic}, while Irving and colleagues have explored mixed linear-nonlinear coupled differential equations using multivariate discrete time series sequences~\cite{irving1997determining}. Jin et al. have demonstrated optimal consumption and portfolio strategies in continuous-time finance models~\cite{jin1997existence}, and Ganesh et al. have analyzed fractional order impulsive stochastic differential equations for controllability~\cite{priya2016controllability}. Additionally, various economic models, including robust economic models and national economy models, have been proposed and analyzed by researchers such as Decardi-Nelson~\cite{decardi2021robust} and Adomian~\cite{adomian1984modeling}, employing decomposition methods and numerical integration techniques. Advancements in computer technology have also led to studies on modeling large power plants~\cite{maffezoni1984computer}. 

The classic Black-Scholes (BS) model stands as a pivotal achievement in financial mathematics, capturing the evolution of market prices for financial assets over time, such as stock options~\cite{black1973pricing}. This model operates under certain assumptions, where variables like asset price \( s \), drift parameter \( \mu \), and volatility \(\sigma \) are treated as constants. Additionally, it assumes frictionless, arbitrage-free, and efficient markets~\cite{gonzalez2016nonlinear},~\cite{edeki2016modified}. However, to address the limitations posed by these assumptions, various modifications have been proposed, including stochastic interest models, jump-diffusion models, stochastic volatility models, and models incorporating transaction costs. Consequently, the Ivancevic option pricing model (IOPM), an alternative to the Black-Scholes equation, has garnered increasing attention in recent years due to its effectiveness in estimating option values~\cite{chen2022soliton},~\cite{ali2023physical},~\cite{elmandouh2022integrability},~\cite{chen2021dark},~\cite{gonzalez2017solving}. The price function \(s = s(t)\) for \(0 \leq t \leq T\) and follows the geometric brownian motion (BM) \(ds = s(\mu dt + \sigma dW_t)\), where \(\mu\) is the drift rate, \(W_t\equiv W(t)\) is the Weiner process. The BS model is given by

\begin{equation}
    \frac{\partial\psi}{\partial t} + rs\frac{\partial\psi}{\partial s} + \frac{1}{2}\sigma^2s^2\frac{\partial^2\psi}{\partial s^2} - r\psi = 0.
\label{Eq1}
\end{equation}
Even though there exists some versions of the BS model for the american stock regulations, the above BS model works for the european stock regulations, where the option is exercised at \(T\). Recently, Vukovic~\cite{vukovic2015interconnectedness} established a connection between two equations: the Schrödinger equation (SE) and the Black-Scholes (BS) model equation, using principles from quantum physics, particularly the Hamiltonian operator. It was further noted that the BS equation could be derived from the SE by employing tools derived from quantum mechanics~\cite{contreras2010quantum}. While the SE is a complex-valued equation, the BS model represents a real-valued equation governing the price function. Eq.(\ref{Eq1}) of the BS model can be extended to one-dimensional option models associated with \(\psi\) and \(s\). As discussed in~\cite{voit2003statistical}, one can determine the probability density function (PDF) by solving the Fokker–Planck equation using classical Kolmogorov probability methods, instead of relying on the option value obtained from the BS equation.

Hence, in the contemporary era, delving into the intricate aspects of economic and financial issues through mathematical modeling stands as a highly explored domain owing to its broad utilization within nonlinear science. These mathematical models typically manifest as nonlinear partial differential equations (NPDEs). Among the extensively researched NPDEs lies in the Ivancevic option pricing model. Ivancevic~\cite{ivancevic2010adaptive} utilized quantum-probabilistic principles to derive a PDF equivalent for the value of a stock option, by constructing a complex-valued function. This approach led to the proposal of a nonlinear model~\cite{cont2001empirical}. Subsequently, this model was rebranded as the Ivancevic Option Pricing Model (IOPM), outlined as follows

\begin{equation}
    i\frac{\partial\psi}{\partial t} + \frac{1}{2}\sigma^2\frac{\partial^2\psi}{\partial s^2} + \beta|\psi|^2\psi = 0,
\end{equation}
where \(\psi(s,t)\) represents the option price function at time \(t\), \(\sigma\) is the volatility which is constant for this model and the adaptive market potential given by \(\beta\) which is the Landau coefficient representing the adaptive market potential. In its simplest from it is the same as the interest rate \(r\) but in this case it depends on the set of adjustable parameters \(\{{w_i}\}\) called the synaptic weight vector components. The adaptive market potential can be related to the market temperature which obeys the Maxwell-Boltzman distribution~\cite{kleinert2009path}. Additional relevant studies have been referenced here to enhance the comprehension of the current analysis. Edeki et al.~\cite{edeki2017analytical} and Gonzalez-Gaxiola and Ruiz de Chavez~\cite{gonz2015solving} examined the Ivancevic option pricing model employing analytical approaches, specifically the projected differential transformation method and a hybrid technique combining the Adomian decomposition method with the Elsaki transform, respectively. Chen et al.~\cite{chen2019new} introduced a novel operator splitting method for resolving the fractional Black-Scholes model concerning American options.

Ivancevic also proposed a full bidirectional quantum neural computation model for option pricing, which is a system of two coupled NLS equations~\cite{ivancevic2010adaptive}. This model describes the option price coupled with the volatility, i.e., the volatility is a function of the asset price \(s\) and time \(t\). Both the option price and the volatility evolves in a common self-organizing market heat potential, therefore representing an adaptive controlled Brownian behaviour of a financial market. 

The time-fractional Ivancevic option pricing model was introduced by Jena et al.~\cite{jena2020novel}. It was introduced due to the fact of the advantages of fractional calculus over traditional integer-order calculus. It has been shown that many real-world phenomena are inadequately described in the realm of integer-order calculus and fractional calculus do better as a tool for solving such problems. Such problems including anomalous diffusion and transport processes~\cite{t1}, viscoelastic materials~\cite{t2},~\cite{t3}, electrochemical processes~\cite{t4}, control systems~\cite{t5}, signal processing~\cite{t6},~\cite{t7} and turbulence and fluid dynamics~\cite{t8},~\cite{t9}.

The purpose of this paper is to study these two models via Modulational Instability analysis and comparing the results.

The dynamics of spatial pattern formation in nonlinear media holds significance across various physical phenomena~\cite{MI1},~\cite{MI2},~\cite{MI3},~\cite{MI4},~\cite{MI5},~\cite{MI6},~\cite{MI7},~\cite{MI8},~\cite{MI9},~\cite{MI10}. Modulational instability (MI) serves as a fundamental mechanism for elucidating pattern formation within a uniform medium. MI arises when a constant-wave background becomes unstable to induced sinusoidal modulations due to the combined influences of nonlinearity and diffraction or dispersion in a spatially nonlinear field. This instability leads to the fragmentation of a uniform medium into pulsed "solitary waves"~\cite{MI2}. Since the Ivancevic option pricing model is a form of a Nonlinear Schrodinger equation (NLS) it can be subjected to this analysis. The MI of a generic NLS excists for a positive product of the interaction and dispersion term, that is they should have the same sign, but since the Ivancevic option pricing model's dispersion term is always positive, instability occurs for positive values of the adaptive market potential (hot adaptive market temperatures). MI for the Ivancevic option pricing model was first performed by~\cite{chen2022soliton}. MI has been used to study several branches of physics, such as biophysics~\cite{BP1},~\cite{BP2},~\cite{BP3},~\cite{BP4},~\cite{BP5}, hydrodynamics~\cite{hydro1},~\cite{hydro2}, optical fibers~\cite{opti1},~\cite{opti2},~\cite{opti3},~\cite{opti4}, Bose-einstein condensates~\cite{BEC1},~\cite{BEC2},~\cite{BEC3},~\cite{BEC4},~\cite{BEC5},~\cite{BEC6}, plasma physics~\cite{pp1},~\cite{pp2},~\cite{pp3},~\cite{pp4},~\cite{pp5},~\cite{pp6}, electrical transmission lines~\cite{et1},~\cite{et2},~\cite{et3}, metamaterials~\cite{mm1},~\cite{mm2},~\cite{mm3},~\cite{mm4},~\cite{mm5},~\cite{mm6},~\cite{mm7}, and finally quantum finance~\cite{chen2022soliton} amongst many others.

The remaining section are as follows: Section \ref{sec5} will focus on the time-fractional Ivancevic option pricing model. Section \ref{sec6} will focus on the linear stability analysis of the model deriving the Bogoliubov dispersion relation of the time-fractional model. Section \ref{sec7} gives the numerical simulations of the time-fractional system and the paper is concluded by Section \ref{sec8}.

\section{The option price coupled Ivancevic model}
\label{sec2}
We investigate the volatility and option price coupled Ivancevic model given by

\begin{align}
i\frac{\partial\sigma}{\partial t} +\frac{1}{2}\frac{\partial^2\sigma}{\partial s^2} + \beta(|\sigma|^2 + |\psi|^2)\sigma =0,
\label{Eq3}
\end{align}
\begin{align}
i\frac{\partial\psi}{\partial t} +\frac{1}{2}\frac{\partial^2\psi}{\partial s^2} + \beta(|\sigma|^2 + |\psi|^2)\psi =0,
\label{Eq4}
\end{align}

where \(\sigma\equiv\sigma(s,t)\) is the volatility, \(\psi\equiv\psi(s,t)\) is the option price and 

\begin{align}
    \beta(r,\omega) &= r\sum_{i=1}^{N}\omega_ig_i,
\end{align}
for \(\dot{\omega_i} = -\omega_i + c|\sigma|g_i|\psi|\)\cite{ivancevic2010adaptive}, is the adaptive market potential. The  \(w-\)ODE is the coupling based continuous Hebbian learning, where \(c\) is the learning rate. The dot product of the synaptic weight vector \(w_i\) and the Gaussian kernel vector \(g_i\) are proportional to the adaptive market-heat-potential. The product can be negative, positive or zero depending on the value of the synaptic weight vector, since the Gaussian kernel vector can neither be zero nor negative. The weight can be dynamically adjust based on various factors, including historical data, market conditions, and external events, affecting the overall behavior and dynamics of the market.

The bidirectional associative memory model Eq.(\ref{Eq3}) - (\ref{Eq4}) effectively engages in quantum neural computation, offering a spatial and temporal quantum extension of Kosko’s bidirectional associative memory (BAM) neural network family\cite{kosko1988bidirectional}. Furthermore, the characteristics of shock waves and solitary waves in the coupled NLS equations might capture phenomena akin to those observed in financial markets, such as the propagation of volatility/prices, and the interaction of shock and solitary waves.

The coupled NLS-system, when \(w-\)learning is absent (i.e., for a constant \(\beta = r\), representing the interest rate), essentially defines the well-established Manakov system, as established by S. Manakov in 1973\cite{manakov1974complete}. This system is proven to be completely integrable due to the existence of an infinite number of involutive integrals of motion. It features solutions characterized as 'bright' and 'dark' solitons. The Manakov system finds application in describing the interaction between wave packets in dispersive conservative media, as well as the interaction between orthogonally polarized components in nonlinear optical fibers, as seen in various studies and references.

\section{Modulational Instability}
\label{sec3}
We perform Modulational Instability on the coupled volatility and option price model. The stationary solution for both the volatility and the option price 
\begin{align}
    \sigma &= \sqrt{n_{\sigma}}e^{-ia_0t},\\
    \psi &= \sqrt{n}e^{-ia_0t},
\end{align}
where \(a_0 = -\beta{(n_{\sigma} + n)}\) and \(n,n_{\sigma}\) are the price and volatility density, respectively. We perturb solution 

\begin{align}
    \sigma = (\sqrt{n_{\sigma}} + \eta)e^{-ia_0t},
\label{Eq8}   
\end{align}

\begin{align}
\psi &= (\sqrt{n} + \chi)e^{-ia_0t}.
\label{Eq9}
\end{align}

Substituting Eqs.(\ref{Eq8}) and (\ref{Eq9}) into Eq.(\ref{Eq3}) and Eq.(\ref{Eq4}) respectively gives the linearized equation for the perturbations \(\eta\) and \(\chi\)

\begin{align}
    i\frac{\partial\eta}{\partial t} &= -\frac{1}{2}\frac{\partial^2\eta}{\partial s^2} - \beta n_{\sigma}(\eta + \eta^*) - \beta\sqrt{n_{\sigma}n}(\chi + \chi^*),\\
    i\frac{\partial\chi}{\partial t} &= -\frac{1}{2}\frac{\partial^2\chi}{\partial s^2} - \beta n(\chi + \chi^*) - \beta\sqrt{nn_{\sigma}}(\eta + \eta^*).
\end{align}
We assume the ansatz \(\eta = \alpha_1e^{i(bs-at)} + \alpha_2e^{-i(bs-at)}\) and \(\chi = \alpha_3e^{i(bs-at)} + \alpha_4e^{-i(bs-at)}\), for real amplitudes \(\alpha_i\),\(i=1,2,3,4\), wavenumber \(s\) and frequency \(a\). Substituting the ansatz and separating the \(e^{i(bs-at)}\) and the \(e^{-i(bs-at)}\) terms, we get a system of equations \(M\vec{\alpha} = \vec{0}\), where \(\vec{\alpha}\) is a vector, \((\alpha_{1}, \alpha{2}, \alpha_3, \alpha_4)^{T}\) \(M\) is a matrix

\begin{widetext}
\begin{equation}
    M = \begin{pmatrix}
        (a + \beta n_{\sigma} - \frac{b^2}{2})&\beta n_{\sigma}&\beta\sqrt{n_{\sigma}n}&\beta\sqrt{n_{\sigma}n}\\
        -\beta n_{\sigma}&(a - \beta n_{\sigma} + \frac{b^2}{2})&-\beta\sqrt{n_{\sigma}n}&-\beta\sqrt{n_{\sigma}n}\\
        \beta\sqrt{nn_{\sigma}}&\beta\sqrt{nn_{\sigma}}&(a + \beta n - \frac{b^2}{2})&\beta n\\
        -\beta\sqrt{nn_{\sigma}}&-\beta\sqrt{nn_{\sigma}}&-\beta n &(a - \beta n + \frac{b^2}{2})
    \end{pmatrix}.
\end{equation}
\end{widetext}
The nontrivial solution of the above matrix are given by equating it's determinant to zero, then solving for the fraquency \(a\), to get the dispersion relation

\begin{align}
a_{1,2} = \sqrt{-T_2\pm\sqrt{T_2^2-T_0}},
\label{Eq13}
\end{align}

\begin{align}
a_{3,4} = -\sqrt{-T_2\pm\sqrt{T_2^2-T_0}},
\label{Eq14}
\end{align}

where 
\begin{align}
\begin{split}
T_0 =\frac{b^8}{16} - \beta\frac{b^6}{4}\left(n + n_{\sigma}\right),
\end{split}
\end{align}

\begin{align}
\begin{split}
T_2 = \beta \frac{b^2}{2}\left(n+n_{\sigma}\right) - \frac{b^4}{4}.
\end{split}
\end{align}

\begin{figure}[h]
    \centering
    \begin{minipage}{0.49\textwidth}
        \includegraphics[width=1\textwidth]{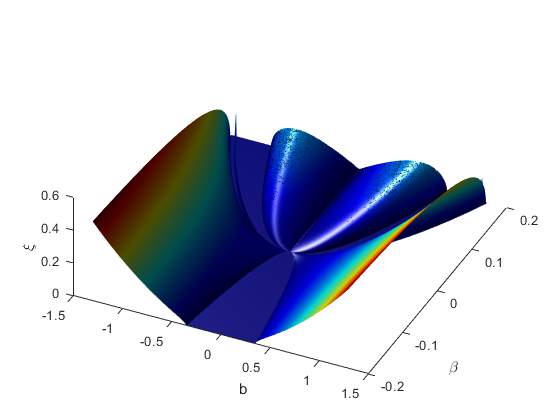}
        \label{fig:2.0}
    \end{minipage}
    \hfill
    \begin{minipage}{0.49\textwidth}
        \includegraphics[width=1\textwidth]{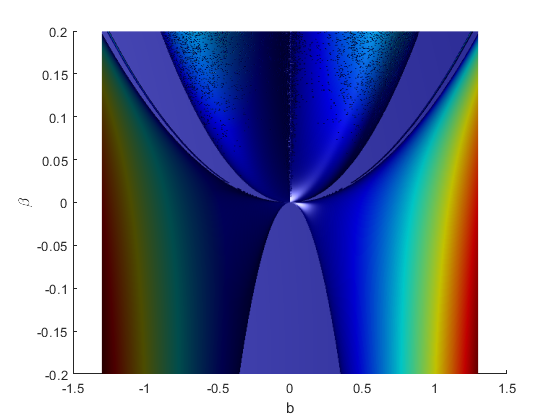}
        \label{fig:2.1}
    \end{minipage}
    \caption{The panels show the variation of the MI gain in the $(b-\beta)$ plane.}
    \label{fig:19972019}
\end{figure}

The above equations, Eq.(\ref{Eq13}) and Eq.(\ref{Eq14}) can be negative or positive, real or complex. Instability occurs when \(a_{i,j}\) is imaginary, that is when the solution of the model are solitons or solitary waves, but for real \(a_{i,j}\), the model assumes plane wave solutions. Hence \(\xi=Im(a)\) is called the MI gain of the system shown in figure \ref{fig:19972019}, depending on the wavelength and the adaptive market potential.

\begin{figure}[h]
    \centering
    \includegraphics[width=\linewidth]{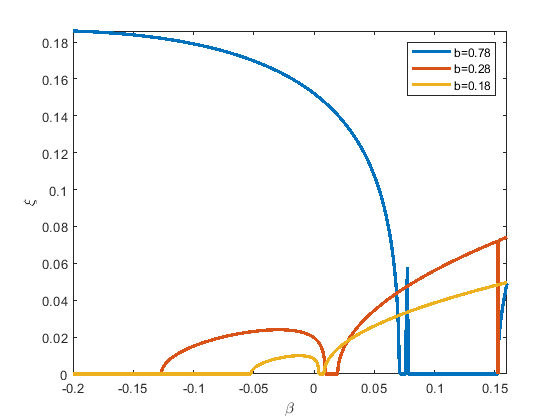}
    \caption{The Figure shows the distribution of MI gain for different values of $b$ (the wavenumber).}
    \label{fig:1.2013}
\end{figure}

The only coupling term in the system of equations is the adaptive market potential.That means that instability conditions are the same for both the volatility and the option price (miscible)\cite{tamilthiruvalluvar2019impact}. Figure \ref{fig:19972019} and \ref{fig:1.2013} shown the conditions for Instability. Even though the simple Ivancevic option pricing model enterpreted as a simple Nonlinear Schrodinger equation, instability occurs for \(\beta>0\) \cite{dauxois2006physics},\cite{rapti2003modulational} the coupled volatility-option price Ivancevic model is unstable for hot market temperatures \(\beta>0\) and also cold market temperatures \(\beta<0\) depending on the values for \(beta\) and \(b\). Figure \ref{fig:1.2013} shows the MI gain for different selected values of the wavenumber \(b\) which is the same for the negative side of the same numbers since the MI gain plot is symetric with respect to the wavenumber axis.

Figure \ref{fig:1.2013}, shows the MI gain for different values of \(b\). For \(b=0.78\) the system is unstable for positive \(\beta\), for \(b=0.28\) and \(b=0.18\) the system is unstable for positive \(\beta\). 
 
\section{Numerical Experiment}

We perform a runge-Kutta method on the coupled volatility option price model. Time in numerical simulations is discrete and increases in timesteps \(t_n + \Delta t\), where at \(t_0\) the assumed ansatz is given by \(\psi(s,0) = \sigma(s,0) = \sqrt(n_{0}) + \epsilon cos(bs)\), where \(\epsilon=0.01\),\(n_0 = n =n_{\sigma} = 10\). In the code we used \(\Delta s = \frac{L}{N}\), where \(L=64\pi\), \(N=2^{10}\) and time step is \(\Delta=\frac{1}{4000}\).

We can see that in this environment the condition give way to solitons, even though we started with a modulated initial condition. In figure the figures below the wave for \(\sigma(s,t)\) and \(\psi(s,t)\) are identical, which is the result of having the same coupling term. In the figures below we observe stable waves at \(\beta = 0.1\) for the \(b=0.78\) case, but for \(\beta = 0.1\) we observe solitons for the \(b=0.28\) and the \(b=0.18\) case. Otherwise we observe solitons for \(\beta = -0.15\) for the \(b=0.78\) case, whereas we observe planewaves for \(\beta=-0.15\) in the \(b=0.28\) and the \(b=0.18\) cases. In all the above cases, the miscibility property of the coupled equations is observable. 

\begin{figure}[h]
    \centering
    \begin{minipage}{0.35\textwidth}
        \includegraphics[width=1\textwidth]{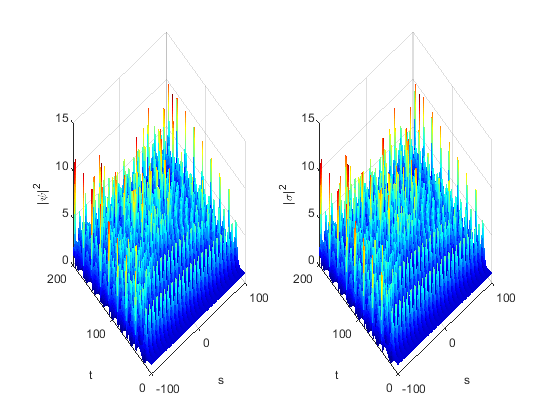}
        \caption{The Figure shows the temporal development of the CW wave of the option price $\psi$ for \(b=0.78\), \(\beta=-0.15\). }
        \label{fig:1.4}
    \end{minipage}
    \hfill
    \begin{minipage}{0.35\textwidth}
        \includegraphics[width=1\textwidth]{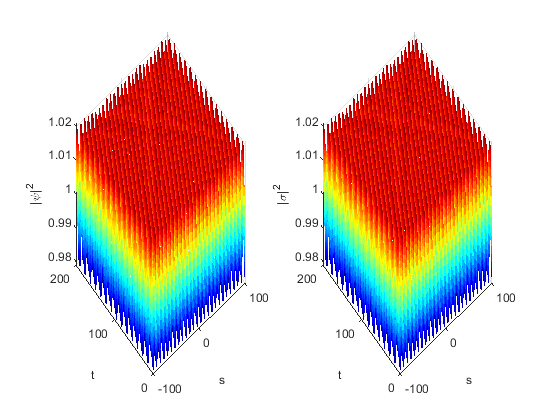}
        \caption{The Figure shows the temporal development of the CW wave of option price $\psi$ for \(b=0.78\), \(\beta=0.1\).}
        \label{fig:1.5}
    \end{minipage}
\end{figure}
\begin{figure}[h]
    \begin{minipage}{0.35\textwidth}
        \includegraphics[width=1\textwidth]{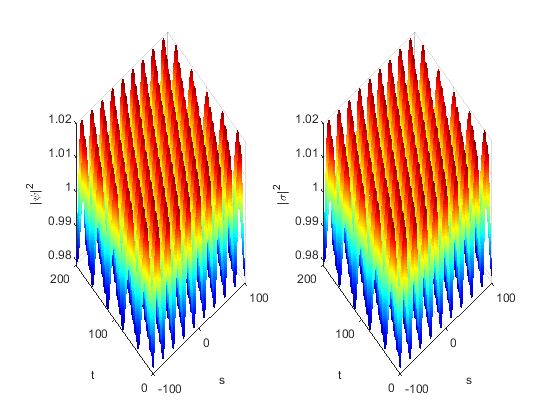}
        \caption{The Figure shows the temporal development of the CW wave of option price $\psi$ for \(b=0.28\), \(\beta=-0.15\).}
        \label{fig:1.6}
    \end{minipage}
    \hfill
    \begin{minipage}{0.35\textwidth}
        \includegraphics[width=1\textwidth]{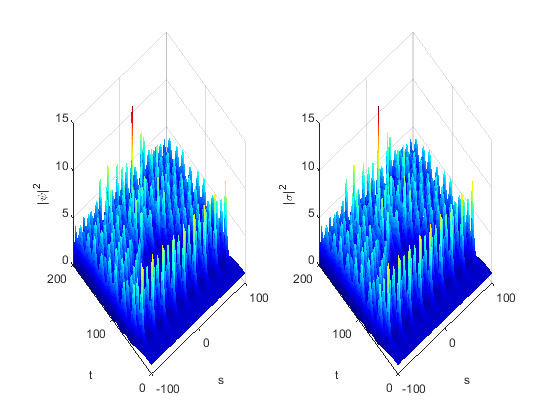}
        \caption{The Figure shows the temporal development of the CW wave of option price $\psi$ for \(b=0.28\), \(\beta=0.1\).}
        \label{fig:1.7}
    \end{minipage}
    \hfill 
    \begin{minipage}{0.35\textwidth}
        \includegraphics[width=1\textwidth]{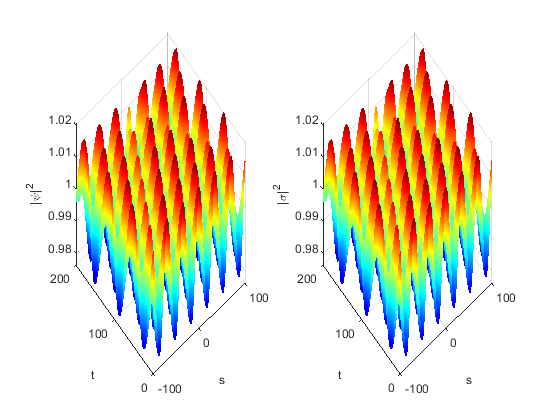}
        \caption{The Figure shows the temporal development of the CW wave of option price $\psi$ for \(b=0.18\), \(\beta=-0.15\).}
        \label{fig:1.7}
    \end{minipage}
    \hfill 
    \begin{minipage}{0.35\textwidth}
        \includegraphics[width=1\textwidth]{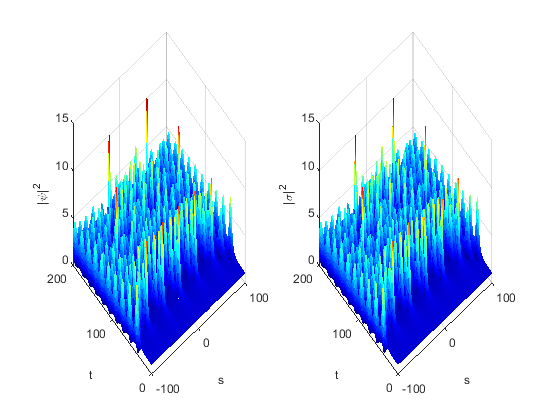}
        \caption{The Figure shows the temporal development of the CW wave of option prices $\psi$ for \(b=0.18\), \(\beta=0.1\).}
        \label{fig:1.7}
    \end{minipage}
\end{figure}
\label{sec4}

\section{The time-fractional Ivancevic Model}
\label{sec5}
Next, we consider the time-fractional Ivancevic model is given by

\begin{equation}
    i\frac{\partial^{\alpha}\psi}{\partial t^{\alpha}} + \frac{1}{2}\sigma^2\frac{\partial^2\psi}{\partial s^2} + \lambda|\psi|^2\psi = 0,
\label{Eq17}    
\end{equation}
where \(0<\alpha<1\) is the Hurst exponent for describing the unpredictability of stock exchange deviations which can be denoted with the aid of the time variation of order \({(dt)^H}\)~\cite{jena2020novel}. Which changes the option pricing model to the non-integer order model.

\section{Linear Stability Analysis}
\label{sec6}
For the stationary solutions of Eq.(\ref{Eq17}) Modulational Instability analysis is studied. A plane wave solution

\begin{equation}
    \psi = \sqrt{n}e^{-ia_0t},
\end{equation}
where \(a_0 = -i(i\lambda n )^{\frac{1}{\alpha}}\).

We pertube the stationary solution

\begin{equation}
    \psi = (\sqrt{n} + \chi(s,t))e^{-ia_0t},
\label{Eq19}
\end{equation}

where

\begin{align}
    \chi(s,t) &= \gamma e^{i(bs - at)} + \delta e^{-i(bs-at)},\label{Eq20}\\
    \chi^*(s,t) &= \gamma e^{-i(bs - at)} + \delta e^{i(bs-at)},
\end{align}
where \(*\) stands for the complex conjugate, and \(b\) and \(a\) are real wavenumber and frequency respectively. Substituting Eq.(\ref{Eq19}) and Eq.(\ref{Eq20}) into Eq.(\ref{Eq17}) we get
\begin{widetext}
\begin{multline}
    i\gamma(i)^{\alpha}(a_0+a)^{\alpha}e^{i(bs-at)} + i\delta(i)^{\alpha}(a_0-a)^{\alpha}e^{-i(bs-at)} - \frac{1}{2}\sigma^2b^2\gamma e^{i(bs-at)}\\-\frac{1}{2}\sigma^2b^2\delta e^{-i(bs-at)}+n\lambda\gamma e^{i(bs-at)}+n\lambda\delta e^{-i(bs-at)} +n\lambda\gamma e^{i(bs-at)} +n\lambda\delta e^{i(bs-at)} \\+ n\lambda\gamma e^{-i(bs-at)} + n\lambda\delta e^{-i(bs-at)} = 0,
\label{Eq22}
\end{multline}
\end{widetext}
by using \(a_0 = -i(i\lambda n)^{\frac{1}{\alpha}}\), and simplifying. Since \(a_0 = -i(i\lambda n)^{\frac{1}{\alpha}} > 1\), that means that \(\frac{a}{a_0} < 1\). Therefore we can use a binomial expansion
\begin{align}
    (a_0 + a)^{\alpha} &= a_0^{\alpha}\left(1 + \alpha\frac{a}{a_0}\right),\label{Eq23}\\
    (a_0 - a)^{\alpha} &= a_0^{\alpha}\left(1 - \alpha\frac{a}{a_0}\right).\label{Eq24}
\end{align}
Substituting Eq.(\ref{Eq23}) and Eq.(\ref{Eq24}) into Eq.(\ref{Eq22}) gives us a system of equations

\begin{equation}
    \begin{pmatrix}
        2n\lambda - \frac{1}{2}\sigma^2b^2-n\lambda\eta_{\alpha} & n\lambda \\ n\lambda & 2n\lambda - \frac{1}{2}\sigma^2b^2+n\lambda\eta_{\alpha}
    \end{pmatrix}
    \begin{pmatrix}
        \gamma\\ \delta
    \end{pmatrix} = \begin{pmatrix}
        0\\0,
    \end{pmatrix}
\end{equation}

where \(\eta_{\alpha} = \left(\frac{ia\alpha}{(in\lambda)^{\frac{1}{\alpha}}} + 1\right)\). By separating the imaginary and real terms. The above's matrix non-trivial solutions can be solved by taking the determinant and equating to zero, and we get

\begin{equation}
    a_{\pm} = \pm\frac{(in\lambda)^{\frac{1}{\alpha}}}{n\lambda\alpha}b\sigma\sqrt{n\lambda - \frac{1}{4}\sigma^2b^2},
\end{equation}
which is the dispersion relation for the time-fractional Ivancevic model. The gain \(\xi\) is the imaginary of the frequency \(Im(a)\) and its the condition for instability, an environment where solitons exists. For \(\alpha = 1\) instability occurs for \(\lambda > 0\) and  \(\frac{\sigma^2b^2}{4n}<\lambda\) which agrees with MI literature, hence stability occurs for \(\lambda < 0\) values and also for \(\frac{\sigma^2b^2}{4n}>\lambda\), which agrees with literature. 

\begin{figure}[h]
    \centering
    \includegraphics[width=\linewidth]{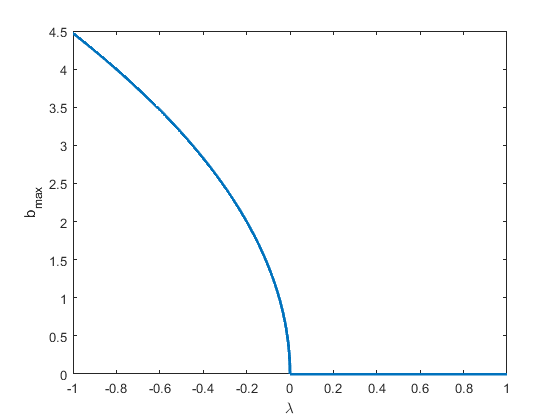}
    \caption{A plot of the real wavelength at maximum gain.}
    \label{fig:2.2}
\end{figure}

\begin{figure}[h]
    \centering
    \includegraphics[width=\linewidth]{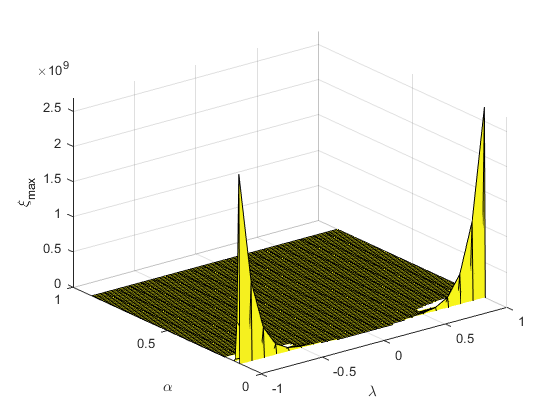}
    \caption{The figure shows a plot of the maximum gain in the \(\alpha-\lambda\) plane}
    \label{fig:2.3}
\end{figure}

Taking a derivative of the dispersion relation and equating it to zero gives us the wavelength at maximum gain figure \ref{fig:2.2}.

\begin{equation}
    b_{max} = \frac{\pm\sqrt{-2n\lambda}}{\sigma},
\end{equation}

and substituting it into the dispertion relation Eq.(17) we get

\begin{equation}
    a_{max} = i\sqrt{3}\frac{(in\lambda)^{\frac{1}{\alpha}}}{\alpha},
\end{equation}
which is the maximum of the dispersion relation. And that's where the maximum gain is comes from, which is \(\xi_{max} = Im(a_{max})\) figure \ref{fig:2.3}.
\begin{figure}[t]
    \centering
    \begin{minipage}{0.49\textwidth}
        \includegraphics[width=1\textwidth]{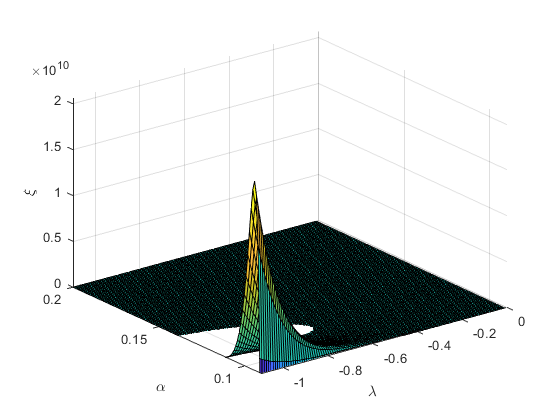}
    \end{minipage}
    \hfill
    \begin{minipage}{0.49\textwidth}
        \includegraphics[width=1\textwidth]{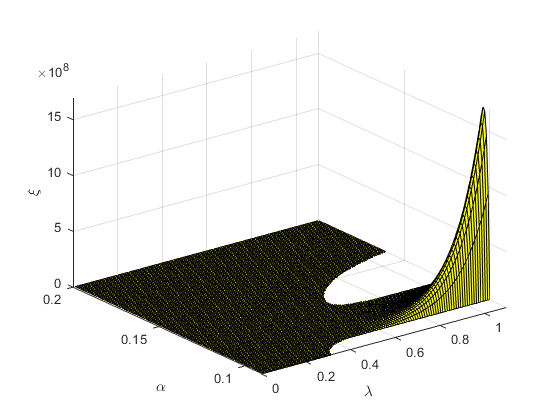}
    \end{minipage}
    \caption{The panels shows the variation of the MI gain in the \(\alpha-\lambda\) plane}
    \label{fig:1.4}
\end{figure}

\begin{figure}[h]
    \centering
    \begin{minipage}{0.49\textwidth}
        \includegraphics[width=1\textwidth]{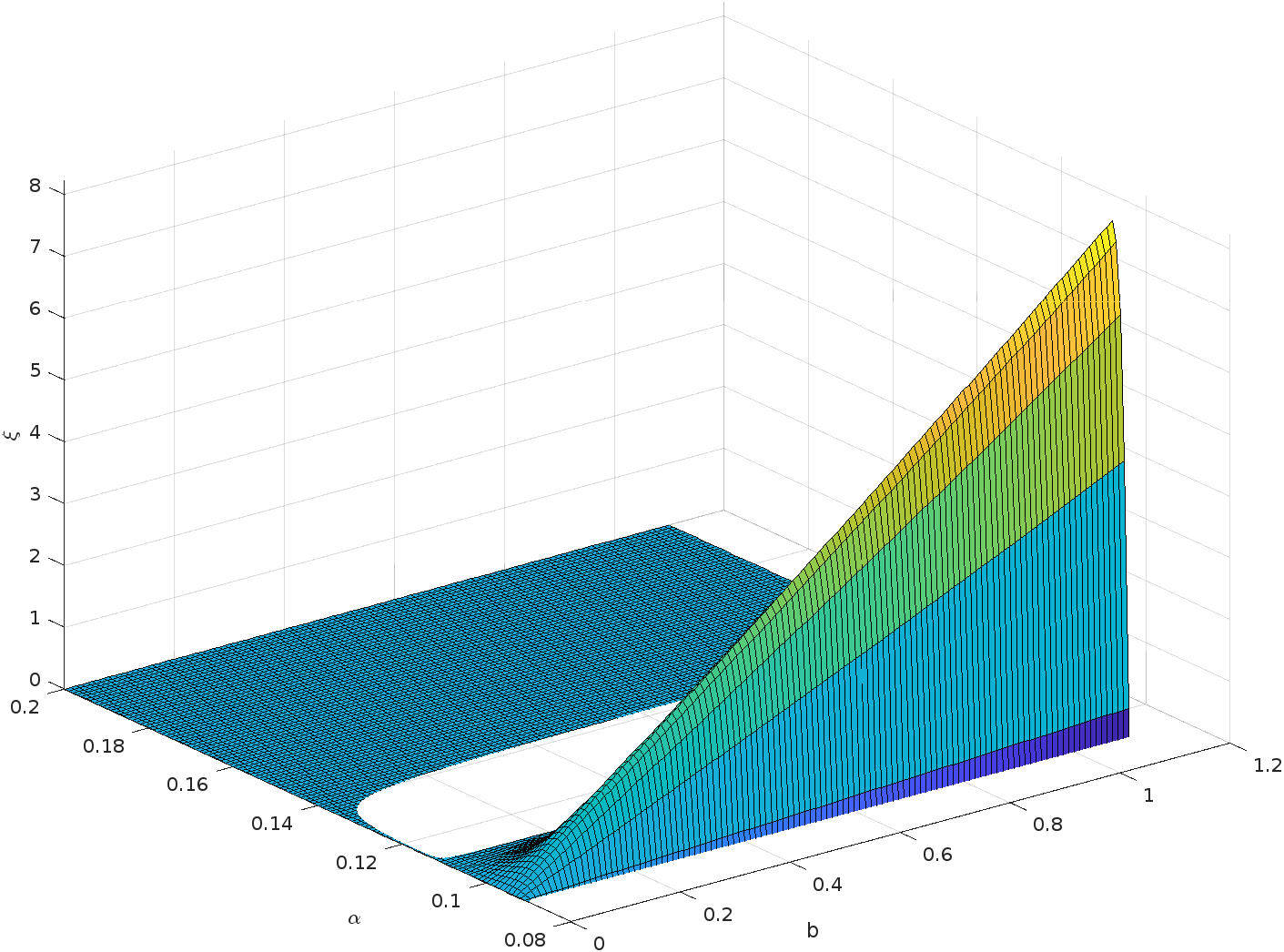}
        \caption{The panels shows the variation of the MI gain in the \(\alpha-b\) plane for \(\lambda=-0.5\)}
        \label{fig:1.5}
    \end{minipage}
    \hfill
    \begin{minipage}{0.49\textwidth}
        \includegraphics[width=1\textwidth]{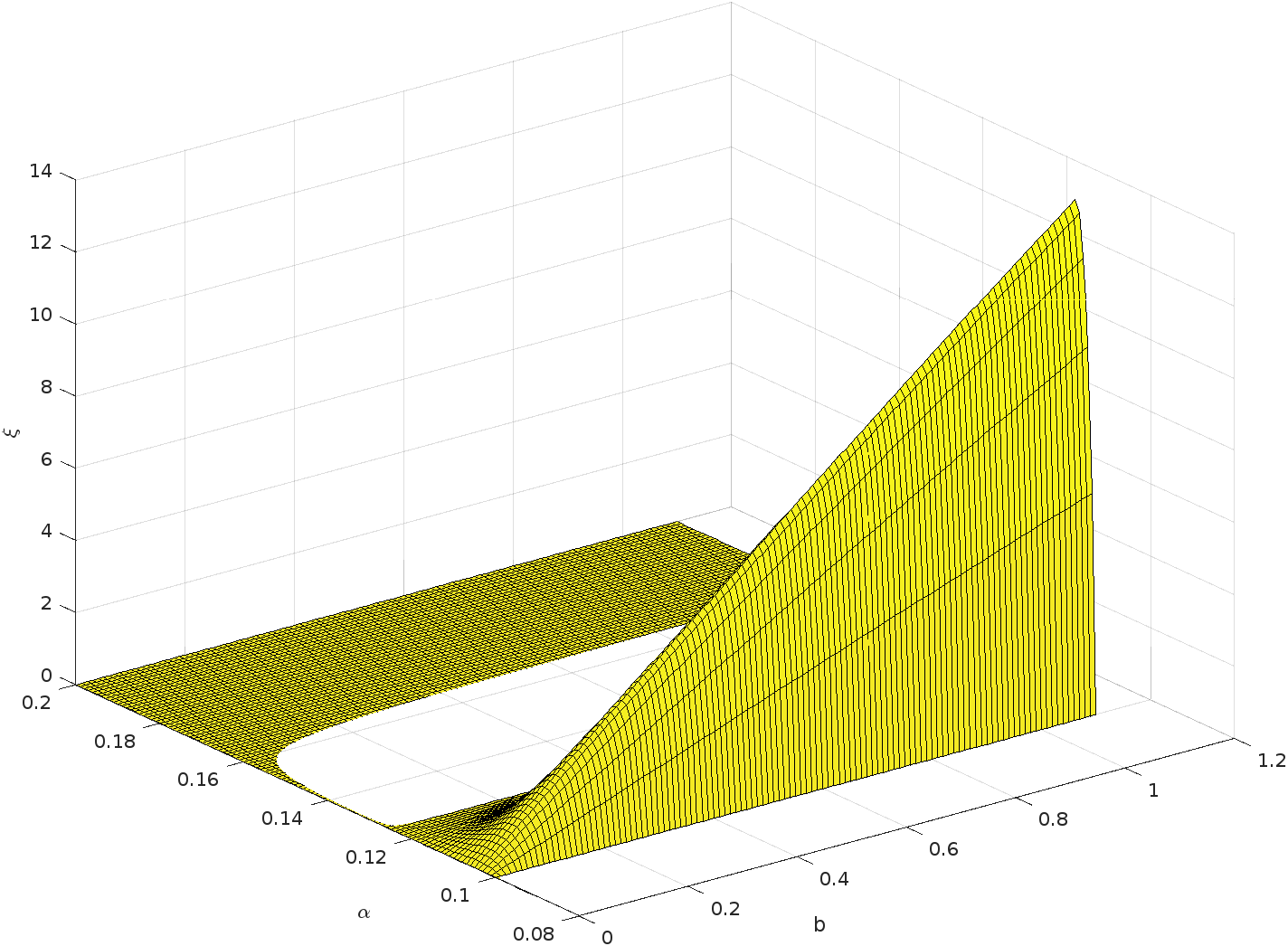}
        \caption{The panels shows the variation of the MI gain in the \(\alpha-b\) plane for \(\lambda=0.5\)}
        \label{fig:1.6}
    \end{minipage}
\end{figure}


\begin{figure}[h]
    \centering
    \begin{minipage}{0.49\textwidth}
        \includegraphics[width=1\textwidth]{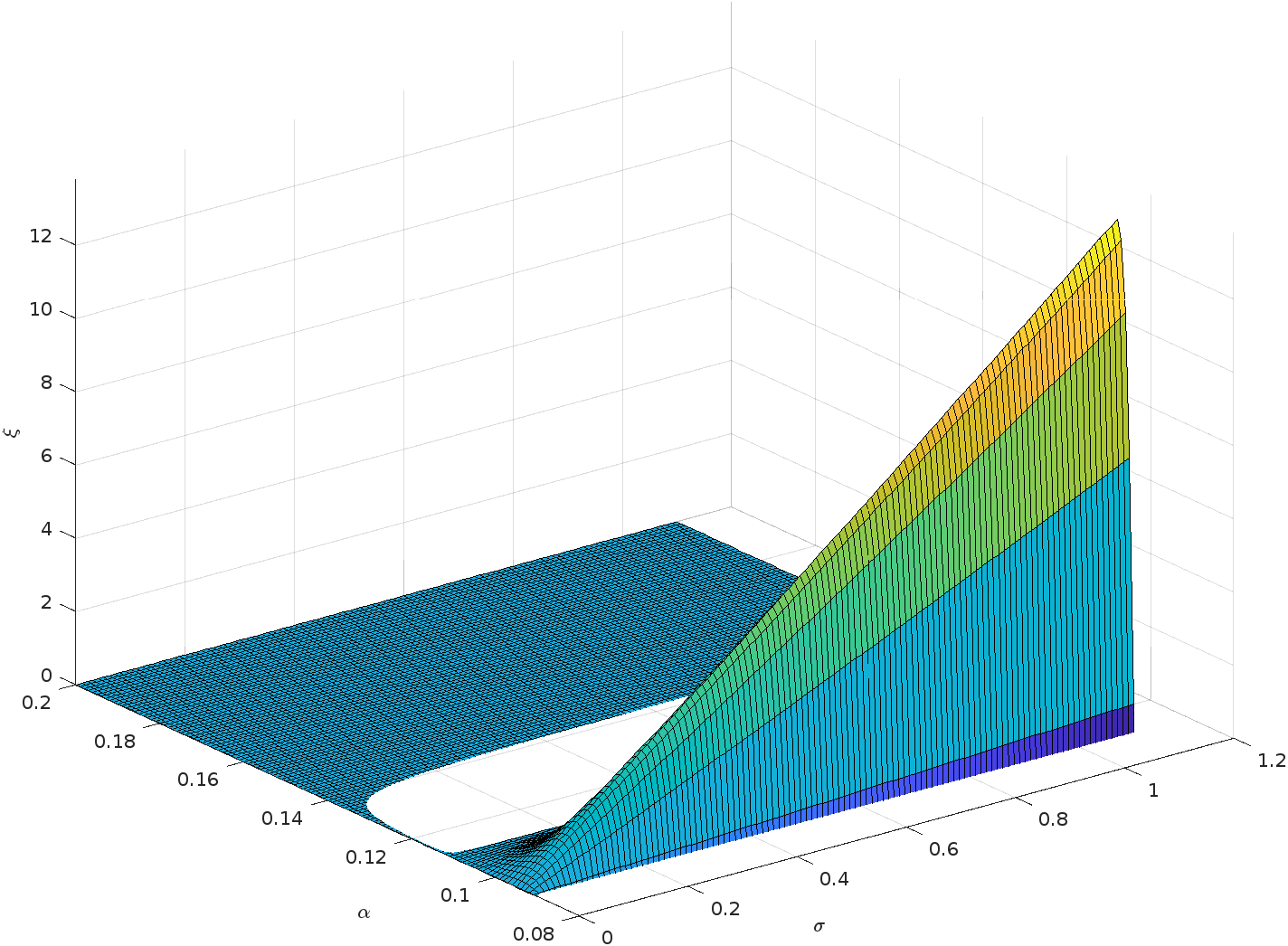}
        \caption{The panels shows the variation of the MI gain in the \(\alpha-\sigma\) plane for for \(\lambda=-0.5\) and \(b=0.5\)}
        \label{fig:2.0}
    \end{minipage}
    \hfill
    \begin{minipage}{0.49\textwidth}
        \includegraphics[width=1\textwidth]{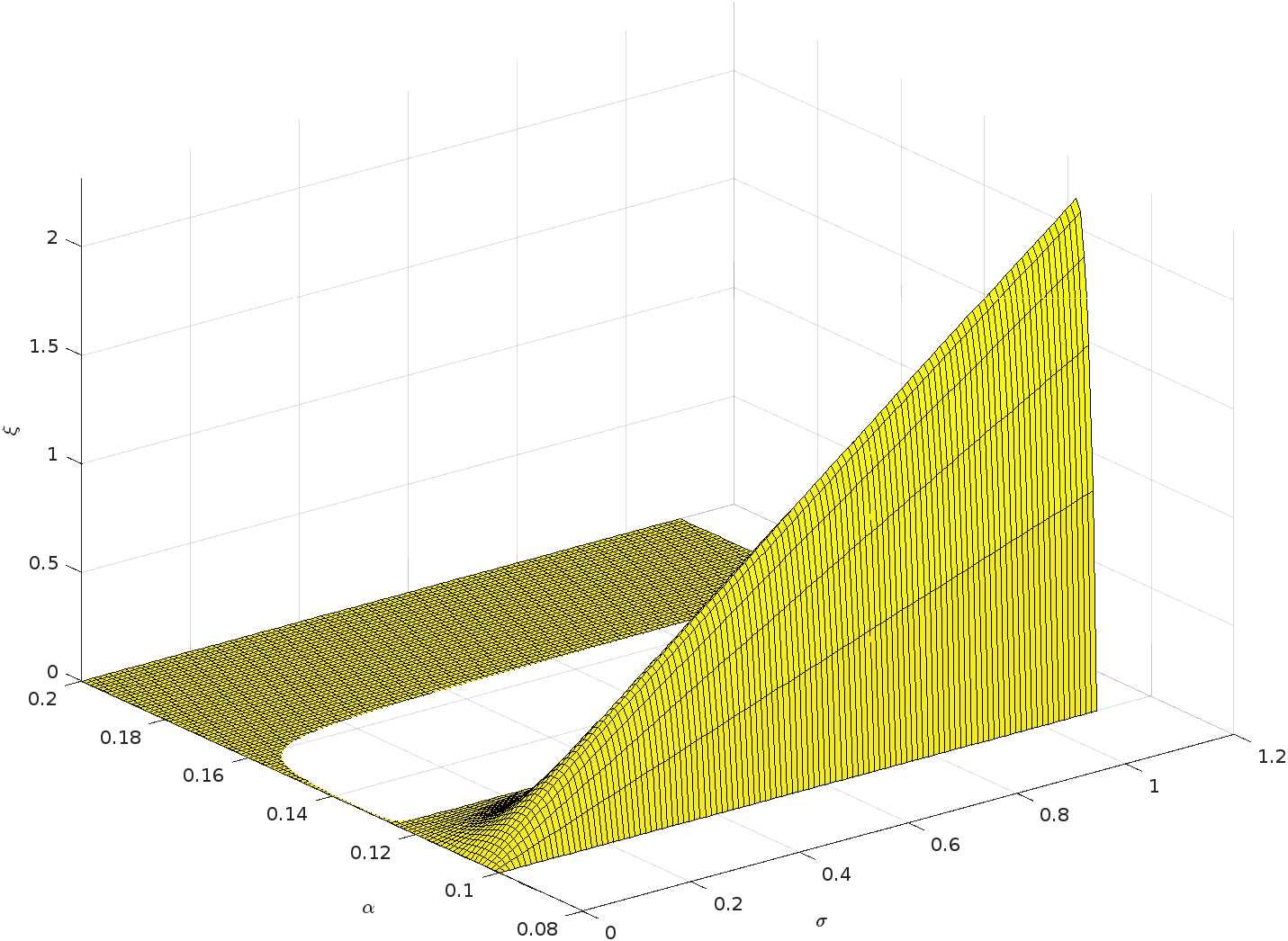}
        \caption{The panels shows the variation of the MI gain in the \(\alpha-\sigma\) plane for for \(\lambda=0.5\) and \(b=0.5\)}
        \label{fig:2.1}
    \end{minipage}
\end{figure}


\begin{figure}[h]
    \centering
    \includegraphics[width=\linewidth]{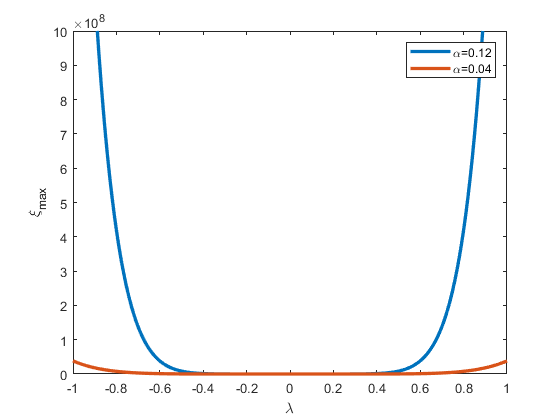}
    \caption{The figure shows a plot of the maximum gain against \(\lambda\) for different values of \(\alpha\)}
    \label{fig:2.4}
\end{figure}

Even though the MI in plots figure \ref{fig:1.4} to figure \ref{fig:2.1} exists for positive values of the adaptive heat market potential and negative values of the adaptive heat market potential depending on other physical conditions, ultimately it is the maximum gain that allows for generation of solitons. Figure \ref{fig:2.3} shows the maximum gain as a function of the Hurst exponential and the adaptive market heat potential. The maximum gain is nonzero for both hot \(\lambda>0\) and cold \(\lambda<0\) market temperatures, but maximum wavelength \(b_{max}\) exists only for cold market temperatures.

\section{Simulations}
\label{sec7}
We perform a 4th order runge-kutta method for our time fractal Ivancevic option pricing model. At time \(t_0\) for discrete time \(t_n + \Delta t\), the solution is \(\psi(s,0) = \sqrt{n} + \epsilon cos(bs)\), where \(\epsilon=0.01\), \(n=10\). 

\begin{figure}[h]
    \centering
    \includegraphics[width=\linewidth]{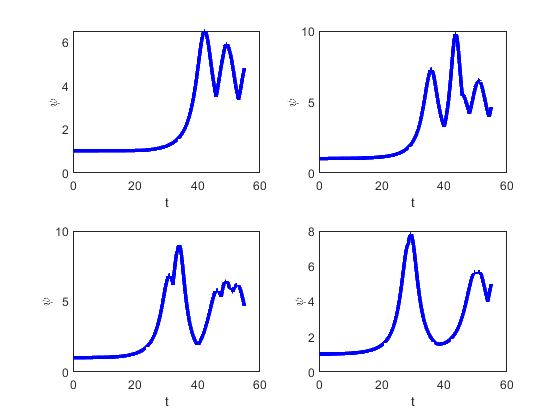}
    \caption{The Figure shows the temporal development of the CW wave of the option prices $\psi$ for \(\alpha=0.12\),\(\lambda=0.2\) with \(b\) increasing from top left to bottom right from \(0.1\), \(0.2\), \(0.3\) to \(0.4\) respectively.}
    \label{fig:2.4}
\end{figure}

The temporal onset of solitons is affected by the wavelength of the wave at \(t_0\). As shown by figure \ref{fig:2.4}.

\section{Conclusion}
\label{sec8}
We have studied both the coupled nonlinear volatility and option price model and the time-fractional option pricing model. In both models instability occurs for both positive and negative values of the adaptive market heat potential, that is both hot and cold market temperatures. For the coupled volatility option price model the instability of the model depends only on the Landau coefficient and the wave number of the perturbed solution Eq.(8) and (9). But for the time-fractional model the Landau coefficient, the Hurst exponential, the volatility, and the wavelength play a part in determining the instability of the model. Direct numerical simulations shows the miscibility of the coupled volatility option price model whereas the numerical simulations of the time-fractional model shows the dependence of the temporal onset of solitons on the wavelength of the initial wave.


\label{sec8}

\end{document}